# *MEASURING THE GRADUALIST APPROACH TO INTERNATIONALIZATION*

Mónica Clavel, Universidad Internacional de La Rioja

Jesús Arteaga-Ortiz, Universidad de Las Palmas de Gran Canaria

Rubén Fernández-Ortiz, Universidad de La Rioja

Pablo Dorta-González, Universidad de Las Palmas de Gran Canaria

***ABSTRACT***

The objective of this paper is to fill a gap in the literature on internationalization, in relation to the absence of objective and measurable performance indicators on the process of how firms sequentially enter external markets. To that end, this research develops a quantitative tool that can be used as a performance indicator of gradualness for firms entering external markets at a sectoral level. The performance indicator is based on firms' export volume, number of years of exporting, geographic areas targeted for export, and when exports were initiated for each area. Additionally, the indicator is tested empirically in the Spanish wine sector. The main contribution of this study is the creation of an international priority index which serves as a valuable and reliable tool because of its potential use in other industry sectors and geographic areas, allowing us to analyze how geographically differentiated internationalization strategies develop.

**KEY WORDS:** international priority index, gradualist approach, psychic distance.

1. **INTRODUCTION**

The study of foreign market entry modes has essentially focused on factors related to the concept of psychic distance (PD) (Martin & Drogendijk, 2014; Zaheer et al., 2012). Numerous factors which influence the attributes of PD have been identified and explain how they determine, at least partially the location choices and sequence of events leading to the entry by firms into new foreign markets (Dikova, 2009). However, the literature does not provide studies which characterize this pattern of behavior. The question which remains unanswered is: how are sequencing decisions made when an industry or sector moves into external markets? In order to answer this question we analyze not the factors which have had an influence, but rather, the steps by which external market entry has been carried out by an industry. By answering this question, this paper fills a research gap by formulating a tool for measuring, independently from any sector, the gradualness of external market entry. We test the tool empirically by applying it to the Spanish wine industry, analyzing the results of their decisions to enter foreign markets throughout its history. The empirical test validates how an objective indicator is generalizable to other industries in order to measure the gradualness pattern of market entry. Having such an indicator will allow us to generalize from the Spanish wine sector by studying the export pattern of gradualness in its opening to foreign markets.

We organize the paper as follows. Section 2 provides the theoretical grounding for the analysis in our paper through a review of the theoretical bases which underpin external market entry mode theory. Next, we identify the different applications of these concepts in the theory and research on International Business. In Section 3 we formulate a set of metrics used in these studies to measure the steps for entering external markets. Section 4 describes the gradualist approach to internationalization. In Sections 5 and 6 we test and verify the validity of the proposed international priority index (IPI) through its application in a key Spanish industry -the wine industry- both from an economic and an exporting perspective. In section 7, we present the main conclusions about the robustness, validity and appropriateness of the proposed IPI, in addition to its implications for research methodology in International Business. In the final section, limitations of the study and directions for future research are provided.

## 2. THEORETICAL FOUNDATIONS

The gradualist approach, or Uppsala model, is considered the best known and most relevant approach to the study of the internationalization of small and medium sized enterprises (Oviatt & McDougall, 1999). The strategy of internationalization is defined as a gradual process based on the seminal study by Johanson and Wiedersheim-Paul developed in 1975. According to this perspective, when a firm considers the development of an internationalization strategy, given its lack of experience and regular information, it will start with sporadic export activities to psychically proximate destinations, incrementally advancing both in the use of resources and the international implications as its international experience increases. In this body of work, it is argued that companies begin their internationalization process in countries that present a closer psychic distance before venturing to more psychically distant countries (O´Grady & Lane, 1996). As argued by Johanson and Vahlne (1990), entering countries that are psychically close reduces the level of market uncertainty that firms face. According to Kogut and Singh (1988), it is easier for these international firms to learn about markets in countries that are psychically close because there is an implicit assumption that psychically close countries are more similar and that similarity is easier for firms to manage than dissimilarity, thereby making it more likely that they will succeed in similar markets (O´Grady & Lane, 1996).

In this way, a firm's first steps will be to psychically close markets since they can be considered extensions of the domestic market, requiring only a small need for adjustment in the operations, systems and processes (Hadley & Wilson, 2003; Johanson & Vahlne, 1977).

However, this gradual and evolutionary process has been criticized (Andersen, 1993; Andersson, 2000; Forsgren, 2001; Hollensen & Arteaga, 2010; McDougall et al., 1994; Reid, 1984) and not empirically tested (Pla & León, 2004). This is due to, among other reasons, the difficulty researchers have in defining an objective, quantitative and measurable indicator which synthesizes and correctly differentiates the method of opening to external markets. Works like Pangarkar (2008), Sahaym & Nam (2013), or Luo et al. (2011) have used indicators like export commitment (external sales over total sales), or export volume, to measure exporting intensity. Other research (Sahaym & Nam, 2013) has measured the exporting experience (number of years exporting), number of foreign customer relationships (Hakansson & Snehota, 2006), destination of the exports or psychic distance in order to study the sequencing processes in internationalization. The latter measure, psychic distance, has been, without doubt, the most widely used in numerous works (Hosseini, 2008) which have tried to model market entry processes. Therefore, in the next section we will discuss how IB literature has addressed the study of psychic distance.

### 3. INDICATORS ON THE ENTRY TO INTERNATIONAL MARKETS: PSYCHIC DISTANCE

As noted by Zaheer et al. (2012), international management is the management of distance. According to Johanson and Vahlne (1977, 1990), PD can be defined as the perceived distance between the home country and a foreign country, resulting in cultural, business, and political differences, i.e. differences in language, political and legal systems, trade practices, industry structure, etc.

The PD concept has been widely used in international business literature (Azar & Drogendijk, 2014; Dikova, 2009) to explain how firms internationalize in terms of market selectivity and how they develop knowledge about foreign markets (Barkema et al., 1996; Benito & Gripsrud, 1992; Dikova, 2009; Johanson & Wiedersheim-Paul, 1975; Klein & Roth, 1990; Kogut & Singh, 1988; Nordström, 1991; Padmanabhan & Cho, 1999; Prime et al., 2009; Rovira & Tolstoy, 2014). Ambos and Hakanson (2014) observed that 29% of the 285 studies on PD analyzed from 1975 to 2011 addressed matters related to market selection, while 25% centered on entry mode related outcomes. The current needs of globalization in destination markets imply that PD is likely to have a greater impact on SMEs' ongoing operations after foreign market entry than it has on market selection decisions (Rovira & Tolstoy, 2014).

To operationalize PD, the following indicators have been used: level of economic development in the importing country; differences in the level of economic development between the exporting countries and the destination countries; level of education in the importing countries; differences in business language, differences in culture and local languages, etc. However, as stated by Langhoff (1997), under this perception of PD it is not possible to conclude that the process of internationalization depends on a company's knowledge, but rather depends on information contained in the public domain.

Barkema et al. (1996), Benito and Gripsrud (1992), Kogut and Singh (1988), and Padmanabhan and Cho (1999), align the meaning of psychic distance more closely to the notion of cultural distance, based on the work developed by Hofstede in 1980, using cultural distance almost as a synonym and proxy for PD (O´Grady & Lane, 1996). Using this line of thought, Ronen and Shenkar (1985), based on a review of empirical works which used Hofstede's index, identified various cultural groups in countries by calculating the difference between the index of the cultural group of the companies' home country, and the different foreign cultural groups, thereby obtaining a measurement of the cultural distance between the different groups of countries and export zones. Finally, Clark and Pugh (2001) measured psychic distance through a construct formed by four independent variables. These were defined as "Market size", "Market influx", "Geographical distance" and "Cultural distance". More recently, following the same line of thought, other authors like Martin and Drogendijk (2014), have proposed measures of PD typically based on publicly available statistics and studies on cultural values, creating a construct formed by cultural distance, physical distance and socioeconomic distance.

Other authors like Klein and Roth (1990) and Nordström (1991), among others, developed a version of the concept of psychic distance through the answers that managers gave to questions on the difficulty of commercializing in foreign markets. This line of thought continues to present difficulties on the globalization and homogeneity of the psychic distance variable. Authors like O´Grady and Lane (1996), affirm that PD should also include industry structure and the competitive environment. As noted by Ambos and Hakanson (2014), without doubt the concept of psychic distance owes much of its attractiveness to its inherent vagueness. Our understanding of the impact of psychic distance has long been constrained by flawed conceptualizations and unreliable measures.

As we have observed, when operationalizing the gradualness construct, an important and significant discordance is found in the literature. The study of PD, or its operationalization, has been criticized (Bae & Salomon, 2010; Drogendijk & Zander, 2010; Nebus & Chai, 2014; Shenkar, 2001, 2012; Shenkar et al., 2008; Smith, 2010; Stöttinger & Schlegelmilch, 2000; Tung & Verbake, 2010; Zaheer et al., 2012) and does not fulfill the needs that, from a research point of view, are required to identify a measure of the gradualist process of internationalization. Thus, if an internationalization pattern is found for a productive area (defining what countries a firm targets and when internationalization occurs), it must take into consideration that each of the companies studied are directing their activities to different areas, in a temporally unequal manner. As they acquire higher levels of international commitment after settling and accumulating experience and knowledge, this knowledge will help them reduce the uncertainty levels (Leonidou & Katsikeas, 1996).

We conclude, therefore, that the empirical literature on International Business needs an indicator that reflects businesses' behavior and foreign market patterns that allows us to quantify for each of them how proximate or distant they are from the pattern of behavior established by the structure of the industry (O´Grady & Lane, 1996). Authors like Davidson (1980) and Clark and Pugh (2001) have partially solved this problem, offering several studies along this line. Based on these studies, in the next section, we propose an objective indicator which, by using market entry date, the entry sequence and the commitment to exporting, will create a measurement of the gradualness process in the internationalization of businesses.

## 4. THE SECTORAL INTERNATIONALIZATION PRIORITY INDEX (IPI) OF AN EXPORT ZONE

Since there is no quantitative measure that indicates if a specific firm, or sector, follows the internationalization gradualist model, we propose the creation of an internationalization priority index that will measure the degree to which firms follow the export behavior outlined in the gradualist methodology literature. The need for this index is supported by the earlier research of Davidson (1980), Clark and Pugh (2001).

We consider three main factors in the export process: the entry order into the export zones that each firm follows; the export width (number of total years since entry) of the firms in each zone; and the export depth (amount of exported product) of the firms in each zone.

Based on Davidson (1980), a comparison by pairs of zones (dyads) in the order followed by each exporting firm allows us to establish a ranking of international priority. Later, this ranking will be weighted by the width and depth of the exports.

For each zone dyad, all the firms that have simultaneously carried out international operations in these two zones are considered, assigning a value of 1 if the firm directed its activity first to the zone whose index we are calculating (zone *z*) and 0 in the opposite case. In this way, we obtain the number of firms that entered zone *z* before entering the rest of the zones considered. This provides an international priority measure for zone *z*. This procedure can be carried out for each zone considered obtaining an international priority ranking for each destination zone.

The export entry year of a firm to each zone provides a measure of the export width. We define the export width of firm *f* in zone *z* in the following way:

$$ExportWidth_z^f = \frac{ExportYears_z^f}{TotalExportYears^f},$$

where $ExportYears_z^f$ is the number of years that firm *f* has been exporting to zone *z,* and $TotalExportYears^f$ is the total number of years that firm *f* has been exporting to any zone. As we are considering exclusively export firms, the denominator of this ratio is always greater than zero.

This ratio belongs to the interval [0.1] and it is a proportion of the years of export to each zone. For example, $ExportWidth_z^f = 0.6$ means that firm *f* has been exporting for 60% of its total export years to zone *z*. In specific cases, a value of 1 means exports to zone *z* began the first year, and a value of 0 means that there are no exports to zone *z*.

The proportion of exports to each zone provides a measure of the depth of exports. We define the export depth of firm *f* in zone *z* in the following way:

$$ExportDepth_z^f = \frac{ExportVolume_z^f}{TotalExportVolume^f},$$

Where $ExportVolume_z^f$ is the amount that firm *f* exports to zone *z,* and $TotalExportVolume^f$ is the total amount that firm *f* exports to all the zones. Notice that the denominator of this ratio is always greater than zero, and $\sum_{z=1}^{n} ExportDepth_z^f = 1$.

This ratio belongs to interval [0.1] and it is a proportion of the exports to each zone. For example, $ExportDepth_z^f = 0.4$ means that firm *f* sends a 40% of its total exports to zone *z*.

Once we have considered all the variables in the export process (the export order, the export width, and the export depth), we define the sectoral *International Priority Index (IPI)* of an export zone *z* as follows:

$$IPI_z = \sum_{\substack{i=1 \\ i \neq z}}^{n} \sum_{f \in F} \left( ExportWidth_z^f \cdot ExportDepth_z^f \right),$$

where *n* is the total number of export zones and *F* is the set of firms that select first *z* then *i*.

This measure can be normalized in intervals [0.1] by dividing by $max_{z=1,2,...,n}\{IPI_z\}$ in order to facilitate its interpretation as a priority rate. Therefore, we define the sectoral *Normalized International Priority Index (NIPI)* of an export zone *z* as following:

$$NIPI_z = \frac{IPI_z}{max_{z=1,2,...,n}\{IPI_z\}}.$$

As an example, *NIPI$_z$=0.85* means that zone *z* has a priority of 85%.

Once the different priority indexes are obtained, we can determine which zones have a greater priority for a given firm or sector. Thus, if we rank the *IPI* obtained for the different zones under consideration, we can establish the order of international priority (sectoral order) in the destination of activities abroad.

Below, in order to clarify the calculation of the Normalized International Priority Index (NIPI), we present an example with four companies who internationalize in four different zones (A, B, C and D).

The first IPI being calculated would be Zone A, which we compare in dyads with the other three zones under consideration. For that purpose, we pick the first dyad (Zone A – Zone B), and observe (see Table 1) how the four companies have developed their international activities in both zones, but only Company 1 and Company 3 entered Zone A before entering Zone B. Thus, using the calculated data on depth and width for Zone A (see Table 2), we add the product of depth and width of the two companies who entered Zone A before Zone B.

Table 1: Start of the internationalization process for each of the companies in the different zones

|        | Year of internationalization ||||  Total Export Years |
|--------|--------|--------|--------|--------|--------|
| Firms  | Zone A | Zone B | Zone C | Zone D | |
| Firm 1 | 1990 | 2000 | 1985 | -    | 28 |
| Firm 2 | 2001 | 1997 | -    | 2005 | 16 |
| Firm 3 | 1986 | 2001 | 1993 | 1980 | 33 |
| Firm 4 | 2005 | 2003 | 1994 | -    | 19 |

Then, we move to the next zone dyad, (Zone A – Zone C). In this dyad, three companies have been directing their export activities to both zones (Company 1, Company 3 and Company 4), and we observe (see Table 1) how only Company 3 entered Zone A before entering Zone C, so we take the product of depth and width from Company 3 in Zone A (see Table 2).

Table 2: Depth and width for each of the companies in the different zones

|        | Zone A || Zone B || Zone C || Zone D ||
|--------|-------|-------|-------|-------|-------|-------|-------|-------|
| Firms  | Depth | Width | Depth | Width | Depth | Width | Depth | Width |
| Firm 1 | 0.30 | 0.82 | 0.20 | 0.46 | 0.50 | 1.00 | -    | -    |
| Firm 2 | 0.20 | 0.75 | 0.40 | 1.00 | -    | -    | 0.40 | 0.50 |
| Firm 3 | 0.10 | 0.82 | 0.40 | 0.36 | 0.20 | 0.61 | 0.30 | 1.00 |
| Firm 4 | 0.50 | 0.42 | 0.30 | 0.53 | 0.20 | 1.00 | -    | -    |

Finally, we compare Zones A and D, where we find that only Company 2 and Company 3 exported to both zones. Of these two companies, Company 2 exported first to Zone A, and Company 3 exported first to Zone D, so we take the product of depth and width of Company 2 in Zone A.

Once the results for the dyadic comparisons of Zone A are obtained, we add them up to obtain the IPI value for Zone A (IPI Zone A).To calculate the IPI of the remaining zones we proceed in the same manner, obtaining the results shown in Table 3.

Table 3: Calculation of IPI for the four zones

| Zones | Zone Dyads | IPI disaggregated | IPI Total |
|---|---|---|---|
| Zone A | A - B | 0.33 | 0.56 |
| | A - C | 0.08 | |
| | A - D | 0.15 | |
| Zone B | B - A | 0.56 | 0.96 |
| | B - C | 0.00 | |
| | B - D | 0.40 | |
| Zone C | C - A | 0.70 | 1.52 |
| | C - B | 0.82 | |
| | C - D | 0.00 | |
| Zone D | D - A | 0.30 | 0.90 |
| | D - B | 0.30 | |
| | D - C | 0.30 | |

Once the IPIs of the different zones in the example are calculated, in order to normalize the index, we divide each of them by the maximum value reached in all zones. In the proposed example, it would be the value reached by the IPI in Zone C (1.52) so the different NIPI values for this example would be as shown in Table 4.

Table 4: NIPI and sectoral order

|  | ZoneA | ZoneB | ZoneC | ZoneD |
|---|---|---|---|---|
| IPI | 0.56 | 0.96 | 1.52 | 0.90 |
| NIPI | 0.37 | 0.63 | 1.00 | 0.59 |
| NIPI x 100% | 37% | 63% | 100% | 59% |
| Sectoral order | 4 | 2 | 1 | 3 |

Once the normalized international priority for the zones used in this example are calculated, and the order of internationalization in the activity sector of the four fictitious companies is considered, the NIPI, in order from high to low, would be as seen in Table 4.

The interpretation of the NIPI x 100% row in Table 4 is as follows. For the analyzed sector, the highest priority zone is Zone C, with a 37% (100-63) higher priority than Zone B, a 41% (100-59) higher priority than Zone D, and a 63% (100-37) higher priority than Zone A. In addition, Zone B has a 4% (63-59) higher priority than zone D, and so on.

## 5. METHODOLOGY

At present, wine cultivation in Spain comprises a total of 69 denominations of origin, all coming from autonomous communities, with the exception of Cantabria and Asturias.

To select the sample population used to test the IPI, we used the SABI (*Sistemas de Análisis de Balances Ibéricos or* Analysis System of Iberian Balances) database, choosing those companies assigned to code CNAE *11.02: Wine-making process*, which were also active and had their registered offices in Spain. This resulted in a total sample size of 2760 Spanish wineries.

The following zones were considered for this study: European Union; Rest of Europe, U.S.A. and Canada; Mercosur, Rest of Latin America, Asia, Australia; and Other Destinations. The selection of these regions is justified by their presence among the main export zones of Spanish wine (ICEX, 2014).

The measuring instrument used was a questionnaire which consisted of a total of 18 items; the fieldwork was carried out in 2010. Table 5 shows the variables considered in the study.

Table 5: Variables introduced in the study

| Variable | Estimator |
|---|---|
| Year of creation | Age |
| Starting year of international activities | Total exportyears |
| Year of entrance in EU | ExportyearsEU |
| Year of entrance in USA | ExportyearsUSA |
| Year of entrance in Rest of Europe | ExportyearsrestEU |
| Year of entrance in Mercosur | Exportyears Mercosur |
| Year of entrance in Rest of Latin America | ExportyearsrestLA |
| Year of entrance in Asia | Exportyears Asia |
| Year of entrance in Australia | Exportyears Australia |
| Year of entrance in others | Exportyearsothers |
| Percentage of exports in EU | Exportdepth EU |
| Percentage of exports in USA | Exportdepth USA |
| Percentage of exports in rest of EU | Exportdepthrest EU |
| Percentage of exports in Mercosur | Exportdepth Mercosur |

| Percentage of exports in rest of Latin Ame. | Export depth rest LA |
| Percentage of exports in Asia | Export depth Asia |
| Percentage of exports in Australia | Export depth Australia |
| Percentage of exports in others | Exportdepthothers |

The number of valid questionnaires obtained was 255, which is a response rate of 9.23%. This sample has a confidence level of 95%, so we can consider it adequate for this study.

Finally, in order to justify the reliability of this sample and eliminate the potential non-response bias, we applied the test suggested by Armstrong and Overton (1977) and Martín et al. (2009). In order to do this, we performed a variance analysis between the questionnaire answers obtained at an earlier time and those obtained at a later time, obtaining a p-value above 0.05, determining the non-existence of significant differences in the items of the two groups of questionnaires, which confirms that the data obtained in our study does not present non-response bias, or bias due to conditioned response as a consequence of the data gathering method used.

## 6. RESULTS

The Spanish wine sector, in addition to its importance to the country's external image, has, at present, great importance for the economic value it generates (1% of Spanish GDP), for the population it employs, and for the role it performs in environmental conservation (Fuentes et al., 2011).

This sector has shown a sustained recovery trend which may principally be attributed to the export levels attained during the years studied. Moreover, according to the Spanish Observatory of the Wine Market (*Observatorio Español del Mercado del Vino,* 2014), the industry has shown a growth in sales volume of 10.10% and 6.53% respectively over the 2012 and 2013 harvests.

Regarding international markets, results are encouraging and, as pointed out by a report published by the Spanish Wine Federation (*Federación Española del Vino,* 2011), Spain has become a net wine exporter. This means that this market's growth is outside Spain's borders, where a great number of the current consumers of Spanish wine are located.

Concerning the destination of Spanish wines in the international markets, data shows that Spanish wineries sell primarily in the European Union (64.79% of the exporting wineries) and American (52.57%) markets, leaving Asia (37.94%) and Oceania (7.60%) far behind.

With the wine sector described, we can analyze the international priority order of its external market opening strategy. By doing so, using the normalized international priority index (NIPI) defined in the previous section, we will be able to determine the strategic pattern used by Spanish wineries as they move into external markets.

As mentioned above, for this study, the following geographic areas were considered: European Union, Rest of Europe, U.S.A. and Canada, Mercosur, Rest of Latin America, Asia, Australia, and Other destinations. The geographic dimension is generally measured by looking at the number of countries to which a firm exports its products (George *et* al. 2005) but following Rugman and Verbeke (2004), we argue that when the international scope of the firm is concerned, regions rather than countries are the relevant units of analysis. The selection of the eight zones is based on an adaptation of the geographic divisions used by Vaaler and McNamara (2004) and Cerrato and Piva (2012), and on their identification as the main export zones for Spanish wine (ICEX, 2014), as mentioned above.

Table 6 shows some central tendency indicators and variable indicators required for the development of NIPI, as well as the age and exporting experience, differentiated by zone.

Table 6: Central tendency and variable measures by export zones

| Zone | Variables | Width | Depth | Export Experience | Age |
|---|---|---|---|---|---|
| EU | Mean | 0.982 | 0.640 | 16.5 | 30.7 |
| | S.D | 0.114 | 0.310 | 17.1 | 28.4 |
| Rest of Europe | Mean | 0.870 | 0.176 | 16.0 | 27.6 |
| | S.D | 0.210 | 0.196 | 9.2 | 28.0 |
| USA and Canada | Mean | 0.847 | 0.301 | 16.8 | 28.1 |
| | S.D | 0.229 | 0.254 | 17.7 | 27.5 |
| Mercosur | Mean | 0.696 | 0.918 | 24.5 | 32.8 |
| | S.D | 0.262 | 0.985 | 29.3 | 32.8 |
| Rest of LA | Mean | 0.813 | 0.167 | 22.5 | 30.6 |
| | S.D | 0.242 | 0.192 | 29.3 | 30.1 |
| Asia | Mean | 0.793 | 0.181 | 15.1 | 25.8 |
| | S.D | 0.252 | 0.198 | 9.6 | 24.9 |
| Australia | Mean | 0.713 | 0,063 | 21.2 | 26.6 |
| | S.D | 0.287 | 0.058 | 11.8 | 13.1 |
| Others | Mean | 0.858 | 0.292 | 15.0 | 30.6 |
| | S.D | 0.312 | 0.320 | 8.7 | 20.6 |

The results obtained show that the entry order into the different zones considered was: European Union, U.S.A. and Canada, Rest of Europe, Other destinations, Asia, Rest of Latin America, Mercosur, and Australia. Additionally, EU has a 69% higher priority than U.S.A. and Canada, the latter has a 23% higher priority than the Rest of Europe, Rest of Europe has a 1% higher priority than Others, etc. (see Table 7).

Table 7: NIPI of the Spanish wine sector

| Zone | IPI | NIPI | NIPI x 100% | Order |
|---|---|---|---|---|
| EU | 64.08 | 1.00 | 100% | 1 |
| Rest of Europe | 5.36 | 0.08 | 8% | 3 |
| USA and Canada | 19.59 | 0.31 | 31% | 2 |
| Mercosur | 0.41 | 0.01 | 1% | 7 |
| Rest of LA | 2.85 | 0.04 | 4% | 6 |
| Asia | 4.11 | 0.06 | 6% | 5 |
| Australia | 0.00 | 0.00 | 0% | 8 |
| Others | 4.42 | 0.07 | 7% | 4 |

These results show that, according to the Uppsala theory, companies initially direct their international activity towards nearby countries in which psychic distance is lower, and as they acquire experience, they start exporting to countries that are more psychically distant.

## 7. CONCLUSIONS AND IMPLICATIONS

The literature on the entrance to external markets has almost exclusively focused on the study of psychic distance, identifying the factors which have an impact on this concept. This article shows several research works which have studied these factors, their composition, and how they determine, at least in a partial way, the destinies and sequencing in the external opening. However, few works have addressed the pattern in this external opening.

Therefore, our first academic contribution is the review of the fragmented and limited literature on this field.

The second one is the creation of a quantitative methodology that can be used as a performance indicator of gradualness for firms entering external markets at a sectoral level. Moreover, through the use of a quantitative, measurable, objective and continuous indicator, this tool can be used as a dependent variable of export behavior's pattern, or an independent variable in studies which, for example, define firms' external opening strategy. In this sense, this work focuses on how the entrance to external markets performed by firms has taken place, rather than the factors which have had an influence in the process.

A firm's first steps will be towards psychically close markets since they can be considered extensions of the domestic market, which implies a lower necessity to adjust operations, systems and processes (Hadley & Wilson, 2003; Johanson & Vahlne, 1977). Nevertheless, this gradual and evolutionary character has been criticized (Andersen, 1993; Andersson, 2000; Forsgren, 2001; Hollensen & Arteaga, 2010; McDougall et al., 1994; Reid, 1984) and not empirically endorsed (Pla & León, 2004). Among other reasons, this is due to the difficulties found by researchers when attempting to establish an objective, quantitative, and measurable indicator, with the power to correctly synthesize and discriminate the external opening method. Based on this, this work proposes a way to measure, in any sector, the gradualness in the entrance to external markets through the use of an indicator. From the entrepreneur and economic agents' perspective, the existence of such indicator may simplify the external opening strategy, thus allowing them to define the attractive of the different countries or regions for the different economic sectors. Therefore it constitutes a third implication, in this case, for decision makers.

The main contribution of this article is the development of the IPI (International Priority Index), an objective indicator of gradualness in the entrance to external markets at a sectoral level which was developed based on the volume of exportation, the years of exportation, the geographic areas of destination, and the starting time of the exportations in each region. This indicator can be applied to any sector and it helps analyze how the internationalization strategies for the different geographic areas are developed, measuring the degree to which firms follow the export behavior outlined in the gradualist methodology literature. Furthermore, the International Priority Index is a measurement of the internationalization activity. This economic indicator allows the analysis of business performance and prediction of future performances. Our indicator is factual -without distortion by opinion, personal feelings, or prejudices- (objective), measurable, accurate and consistent so that data collected from different sources over time are not skewed due to the different sources or the different periods (well-defined). Furthermore, our indicator has a direct relationship to the internationalization process (valid) and it is easy to obtain.

In that sense, this article develops a quantitative tool that can be used as a performance indicator of gradualness for firms entering external markets at a sectoral level. Therefore, we consider that this tool is a contribution to the international business literature, and an important attempt to aid in its development.

The tool was empirically tested by applying it to the Spanish wine industry, analyzing the results of their decisions to enter foreign markets throughout its history, and validating the generalizable character of this objective indicator. Having such an indicator allows us to generalize from the Spanish wine sector by studying the export pattern of gradualness in the opening to foreign markets.

Regarding the Spanish wineries, our findings show that, according to the Uppsala theory, companies initially direct their international activity towards nearby countries in which psychic distance is lower, and as they acquire experience, they start exporting to countries that are more psychically distant.

## 8. LIMITATIONS AND FUTURE RESEARCH DIRECTIONS

The limitations of the study should be considered when the results are interpreted. Firstly, although the empirical data focused on a sample of Spanish wineries, the findings could be of interest to firms in other countries. However, the readers should exercise caution in attempting to generalize this study's findings to considerably different economic settings.

Regarding future research directions, and taking into account that the main contribution of this study is the creation of an international priority index which serves as a valuable and reliable tool because of its potential use in other industry sectors and geographic areas, allowing us to analyze how geographically differentiated internationalization strategies develop, it would be interesting to replicate similar studies in distinct geographical contexts, so the results could be generalized to larger populations.

Secondly, this study was centered on a cross sectional research design performed in a given moment in time, with enterprises operating in different export stages or with different years of experience, thus, no longitudinal analysis was performed. Future studies should consider employing longitudinal analysis in order to illustrate the dynamics of exporting. Thirdly, it may also be advisable to carry out similar investigations within various industries, as well as to differentiate the results obtained according to the specific overseas markets served.

Despite these limitations, and although results need to be confirmed by further research, the study did provide preliminary answers to the research goals.